\documentclass[draft]{agujournal2019}
\usepackage{url} 
\usepackage{soul}
\usepackage{amsmath}
\draftfalse

\journalname{JGR: Planets}

\begin{document}

%
%

\title{Effect of impact velocity and angle on deformational heating and post-impact temperature}

%
%




\authors{S. Wakita\affil{1,2}, H. Genda\affil{3}, K. Kurosawa\affil{4}, T. M. Davison \affil{5}, and B. C. Johnson \affil{1,6}}

\affiliation{1}{Department of Earth, Atmospheric, and Planetary Sciences, Purdue University, West Lafayette, IN, USA}
\affiliation{2}{Department of Earth, Atmospheric and Planetary Sciences, Massachusetts Institute of Technology, Cambridge, MA, USA}
\affiliation{3}{Earth-Life Science Institute, Tokyo Institute of Technology, Meguro, Japan}
\affiliation{4}{Planetary Exploration Research Center, Chiba Institute of Technology, Narashino, Japan}
\affiliation{5}{Department of Earth Science and Engineering, Imperial College London, London, UK}
\affiliation{6}{Department of Physics and Astronomy, Purdue University, West Lafayette, IN, USA}





\correspondingauthor{Shigeru Wakita}{shigeru@mit.edu}



\begin{keypoints}
\item We examined the dependence of impact heating on impact angle and velocity using a shock physics code.
\item The amount of heated mass is similar among $>45^\circ$ impacts, while it is smaller for shallower impacts.
\item We derived an empirical formula for the cumulative heated mass over 1000 K during oblique impacts.  
\end{keypoints}

%
%

%
%


\begin{abstract}
The record of impact induced shock-heating in meteorites is an important key for understanding the collisional history of the solar system.
Material strength is important for impact heating, but the effect of impact angle and impact velocity on shear heating remains poorly understood.
Here, we report three-dimensional oblique impact simulations, which confirm the enhanced heating due to material strength and explore the effects of impact angle and impact velocity. 
We find that oblique impacts with an impact angle that is steeper than 45 degree produce a similar amount of heated mass as vertical impacts.
On the other hand, grazing impacts produce less heated mass and smaller heated regions compared to impacts at steeper angles.
We derive an empirical formula of the heated mass, as a function of the impact angle and velocity. 
This formula can be used to estimate the impact conditions (velocity and angle) that had occurred and caused Ar loss in the meteoritic parent bodies.
Furthermore, our results indicate that
grazing impacts at higher impact velocities could generate a similar amount of heated material as vertical impacts at lower velocities.
As the heated material produced by grazing impacts has experienced lower pressure than the material heated by vertical impacts, our results imply that grazing impacts may produce weakly shock-heated meteorites.
\end{abstract}

\section*{Plain Language Summary}
Meteorites are extraterrestrial materials that have been delivered to the Earth from asteroids. 
The materials in meteorites can record information about their formation and subsequent evolution. 
Thus, they are excellent sources of information used to explore the history of the solar system.
One such feature recorded is evidence of shock: high pressures and temperatures caused by collisions between asteroids.
Previous work investigating impacts found that material strength is a key factor in determining the amount of impact heating, especially in low-speed collisions, like those expected to occur in the main asteroid belt. 
In this work, we explore various oblique incidence impacts to study the effects of material strength by using a shock physics model.
We confirm that material strength plays a key role in oblique impacts, just as in head-on impacts. 
Our results show that head-on and 45 degree impacts can generate nearly the same amount of heated mass in total. 
However, more oblique impacts with a shallower angle produce less heated mass than other steeper-angle impacts (i.e., head-on and 45 degree impacts).
We also find that low-speed vertical impacts and high-speed grazing impacts can produce the same amount of material in asteroids that have experienced a given temperature. 

%
%

%


%
%
%
%

\section{Introduction}
Asteroids may impact other terrestrial bodies at various impact velocities and angles.
Current asteroids collide with each other at a mean impact velocity of 5 km/s \cite{Bottke:1994,Farinella:1992},
while asteroids hit the Moon and Mars at over 10 km/s \cite{Ivanov:2001,Yue:2013}. 
We observe the evidence of impacts as craters on Moon, Mars, and asteroids,
even on small-sized bodies, like Ryugu and Bennu \cite{Walsh:2019,Morota:2020}.
As asteroids have experienced such collisions over the history of the solar system,
exploring ancient evidence of impacts can reveal the early collisional history of asteroids \cite<e.g.,>{Sugita:2019}.
Ejecta produced by impacts on asteroids sometimes travel to the Earth and are found as meteorites. 
Shock metamorphism in meteorites provides evidence of past impacts on their parent bodies \cite<e.g.,>{Keil:1994,Scott:2002}.
Using shock-induced textures in meteorites, we categorize them by the degree of shock metamorphism \cite{Stoffler:1991, Scott:1992, Rubin:1997, Stoffler:2018}.
When a strong shock propagates in the parent body, it might lead to the melting of materials. 
Weak shocks, however, are unable to produce melted textures; 
weakly shocked meteorites have experienced shock pressure less than 40 GPa.
Nevertheless, weakly shocked meteorites also show evidence of moderate impact heating ($\sim700-800$ $^\circ$C),
such as the dehydration of phyllosilicate minerals \cite{Nakamura:2005,Nakato:2008,Abreu:2013}
and the reset of $^{40}$Ar-$^{39}$Ar ages \cite{Bogard:1995,Bogard:2011,Weirich:2012,Cohen:2013}.
Thus, studying impact heating is essential for a better understanding of the solar system.

Oblique impacts are more likely to occur than head-on impacts in the solar system \cite{Shoemaker:1962}.
The most probable impact angle ($\theta_{\rm imp}$) is $45^\circ$. 
The probability of impact occuring with an impact angle between $\theta_{\rm imp}$ and $\theta_{\rm imp} + d\theta_{\rm imp}$ is given as $\sin(2\theta_{\rm imp}) d\theta_{\rm imp}$  \cite{Shoemaker:1962}.
To understand the nature of oblique impacts, numerical simulations have been performed 
using smoothed particle hydrodynamics \cite<SPH; e.g.,>{Genda:2012,Monaghan:1992,Benz:1994,Jutzi:2015,Sugiura:2021,Okamoto:2020,Citron:2022}
and grid-based hydrodynamic codes \cite<CTH, SOVA, iSALE-3D; e.g.,>{Pierazzo:2000, Pierazzo:2000a, Elbeshausen:2009,Elbeshausen:2011,Elbeshausen:2013,Davison:2014,Artemieva:2017,Wakita:2019a}. 
Previous work has examined the crater volume and heated mass 
\cite{Pierazzo:2000, Pierazzo:2000a, Elbeshausen:2009, Elbeshausen:2013, Davison:2011, Davison:2014}. 
The volume heated to any given temperature depends on the impactor diameter, mass, velocity, and angle \cite{Pierazzo:2000a}.
Recently, the material strength in the target has been recognized as an additional important parameter for heating \cite{Quintana:2015, Kurosawa:2018, Kurosawa:2021a, Wakita:2019, Wakita:2019a}.

As extraterrestrial materials have experienced impacts on their parent asteroids, impact heating is crucial to understand their record. 
When we consider the material strength in rocks, the temperature increase is much higher than previously expected \cite<e.g.,>{Kurosawa:2018}.
Dissipation of the kinetic energy due to plastic deformation in pressure-strengthened rocks 
is equivalent to an increase in internal energy, leading to higher temperatures during, and after, decompression.
\citeA{Kurosawa:2018} confirmed the post-shock heating due to plastic deformation in head-on impacts using a shock physics code.
Although the following work explored an oblique impact of 45$^\circ$ at 5 km/s \cite{Wakita:2019a}, impacts at a variety of impact angles and velocities must be explored to better understand the effects of deformational heating.
\citeA{Davison:2014} showed the dependence of the heated mass on impact angle, however, the estimated post-shock temperature is based on the peak shock pressure, 
and ignores enhanced deformational heating \cite{Kurosawa:2018}.

Here we perform oblique impact simulations with material strength to examine the dependence of impact heating on impact velocities and angles. 
While we vary impact velocity and angle systematically with the same method as in previous work \cite{Wakita:2019a}, 
we also conduct a companion series of the impacts without material strength. 
Simulations without material strength have no deformational heating.
Thus, comparison of simulations with and without strength allows us to quantitatively determine the effect of deformational heating \cite{Kurosawa:2018, Wakita:2019a}.
Considering 1000 K as a reference temperature, which is the temperature for Ar age resetting \cite<following previous work, e.g.,>{Marchi:2013}, 
we provide an empirical relationship of the heated mass which exceeds that temperature.

\section{Methods}
We use the iSALE-3D shock physics code \cite{Elbeshausen:2009,Elbeshausen:2011,Collins:2016}, to simulate a spherical impactor striking a flat-surface target at oblique angles. 
iSALE-3D uses a solver as described in \citeA{Hirt:1974}.
iSALE-2D is limited to simulating vertical-incidence impacts due to its use of axial symmetry. 
Thus, to simulate a range of impact angles, we use the fully three-dimensional version, iSALE-3D. 
The results of iSALE-2D and vertical impacts of iSALE-3D have previously been shown to agree well \cite<e.g.,>{Elbeshausen:2009, Davison:2011, Davison:2014, Raducan:2022}. 
Note that some examined about crater formation and impact ejecta, but \citeA{Davison:2014} compared the impact heating (similar to ours but without the shear heating).
Since material strength is important for studying the effect of impact heating, we employ a strength model appropriate for rocky materials \cite<see below,>{Collins:2004,Melosh:1992,Ivanov:1997}.  
It is well known that porous targets produce more melt than non-porous targets in iSALE-2D \cite{Wunnemann:2006,Davison:2010}, and
a porosity compaction model is implemented into the iSALE-3D \cite{Wunnemann:2006,Collins:2011}. 
However, to reduce the parameter space and compare it with previous work \cite<e.g.,>{Kurosawa:2018},
we only consider non-porous materials in this study.
Nevertheless, we can apply our results in this study to well-consolidated rocky materials, such as ordinary chondrites having a porosity less than 10\% \cite<e.g.,>{Flynn:2018,Ostrowski:2019}.
We assume the impactor and the target have the same composition of dunite, using the ANEOS equation of state \cite{Benz:1989}.
In this work, we model dunite as non-porous with material strength. 
Since the dunite has a well-defined equation of state \cite{Benz:1989}, it is widely used to simulate the bodies in the inner solar system \cite<e.g.,>{Davison:2010, Johnson:2015}. 
It also represents meteoritic material well \cite<ordinary chondrite,>{Svetsov:2015}. 
Thus, material parameters for dunite without porosity in this work are representative of compact bodies in the inner solar system.
As previously noted, simulations without strength are only used to quantify the effects of material strength and are not meant to represent specific solar system bodies.
For the input parameters, we take the same values shown in \citeA<Table S1 of>{Kurosawa:2018}.

When we simulate impacts with material strength, we use two models to describe the yield strength of intact and damaged rock; 
the Lundborg strength model \cite{Lundborg:1968} and the Drucker-Prager model \cite{Drucker:1952}, respectively.
We combine these two models using a damage parameter $D$ which depends on total plastic strain \cite<e.g.,>{Collins:2004}.
$D$ ranges from 0 (intact rocks) to 1 (thoroughly fractured rock).
A damage parameter $D$ is initially set as 0 and damage is accumulated as material deforms and accumulates plastic strain according to the damage model of \citeA{Ivanov:1997}. 
When the shock pressure exceeds the Hugoniot elastic limit, materials become thoroughly damaged. 
We find material is completely damaged out to approximately 4 impactor radii from the point of impact (see also Fig. 5 of \citeA{Wakita:2019a}).
The yield strength $Y$, that of intact rock $Y_i$, and that of damaged rock $Y_d$ are written as follows,
\begin{equation}
    Y = (1-D)Y_i + DY_d, \label{eq:D}
\end{equation}
\begin{equation}
    Y_i = Y_{\rm coh,i} + \cfrac{\mu_{\rm int}P}{1+\cfrac{\mu_{\rm int}P}{Y_{\rm limit}-Y_{\rm coh, i}}},  \label{eq:Lund}
\end{equation}
\begin{equation}
    Y_d = \rm min (Y_{\rm coh}+\mu P, Y_{\rm i}),  \label{eq:DrPr}
\end{equation}
where $Y_{\rm coh,i}$ is the cohesion for intact rock at zero pressure, 
$\mu_{\rm int}$ is the coefficient of internal friction for intact rock, 
$P$ is the temporal mean pressure,
$Y_{\rm limit}$ is the limiting strength at high pressure,
and $\mu$ is the damaged friction coefficient, respectively.
To simultaneously handle both intact and thoroughly fractured rocks (Equation 1), we use the Lundborg model for intact rock (Equation \ref{eq:Lund}) and the Druker-Prager model for the thoroughly fractured rock (Equation \ref{eq:DrPr}), respectively.
The damaged friction coefficient $\mu$ is one of the important parameters for impact heating  \cite{Kurosawa:2018,Wakita:2019, Wakita:2019a}.
The dependence of the damaged friction coefficient has been explored and the smaller value represents the case without material strength \cite{Kurosawa:2018}. 
Following the fiducial case in their work, we also
adopt $\mu$ = 0.6 which is a typical value pertaining to granular materials made of rocky materials \cite<e.g.,>{Collins:2004}.
Note that the limiting strength $Y_{\rm limit}$ is another influential parameter \cite<see also Figure S1 of>{Kurosawa:2018}.
The limiting strength, also known as the von-Mises strength, is the maximum strength material can have regardless of confining pressure. 
Following \citeA{Kurosawa:2018}, we use $Y_{\rm limit}$ of 3.5 GPa in this work.

To examine the dependence of impact heating on impact properties,
we vary the impact velocity ($v_{\rm imp}$ = 2, 3, 5, 7, 10 km/s) 
and the impact angles ($\theta_{\rm imp}$ = 15$^\circ$, 30$^\circ$, 45$^\circ$, 60$^\circ$, 75$^\circ$, 90$^\circ$). 
Note that we measure the impact angles from the target surface, i.e., 90$^\circ$ is a vertical impact.
We fix a radius of the impactor ($R_{\rm imp}$) as 2 km with a resolution of 0.1 km, which corresponds to 20 cells per projectile radius (CPPR).
In this study, we assume that the size ratio of the impactor to the target is small enough to neglect the curvature of the target body.
A resolution of 20 CPPR is sufficient to resolve cumulative heated mass as demonstrated in previous work \cite<see Text S3 and Figure S2 in>{Wakita:2019a}.
As the numerical diffusion exaggerates the temperature of material near the contact boundary between the impactor and the target \cite{Kurosawa:2018}, we need to omit these artificial overheated regions from further analysis.
In our case, that region corresponds to 0.3 times of impactor mass ($M_{\rm imp})$. 
Though we count that region in the results, we do not discuss the results that are less than 0.3 $M_{\rm imp}$ (see following sections). 
We place Lagrangian tracer particles in each cell of a high-resolution zone at the initial state of simulations. 
Note that the computational region in iSALE-3D consists of a high-resolution zone and an extension zone.
The cell size in the extension zone is larger than that in the high-resolution zone and increases according to the distance from the boundary between the high-resolution zone \cite<see also Figure 1 of>{Davison:2011}. 
To track the highly heated region (i.e., $>$ 1000 K), we take the number of high-resolution cells as 220 in horizontal direction (x), 100 in vertical direction (z), and 150 in depth direction (y), respectively.
We also note that we horizontally shift the origin of the impact point according to the impact angle; the origin is located in the middle at  $\theta_{\rm imp}$ = 90$^\circ$ and it is shifted 20 cells in the downrange direction at $\theta_{\rm imp}$ = 45$^\circ$.
The total number of tracer particles is as large as $\sim 2 \times 10^6$.
The Lagrangian tracers move through the Eulerian grid tracking the temperature and movement of a parcel of material. 
Note that we neglect radiative and conductive cooling during impact events. 
These are less effective than the cooling due to expansion over the timescales considered here \cite<e.g.,>{Sugita:2002}.

\section{Results}
\begin{figure}
\noindent\includegraphics[width=0.9\textwidth]{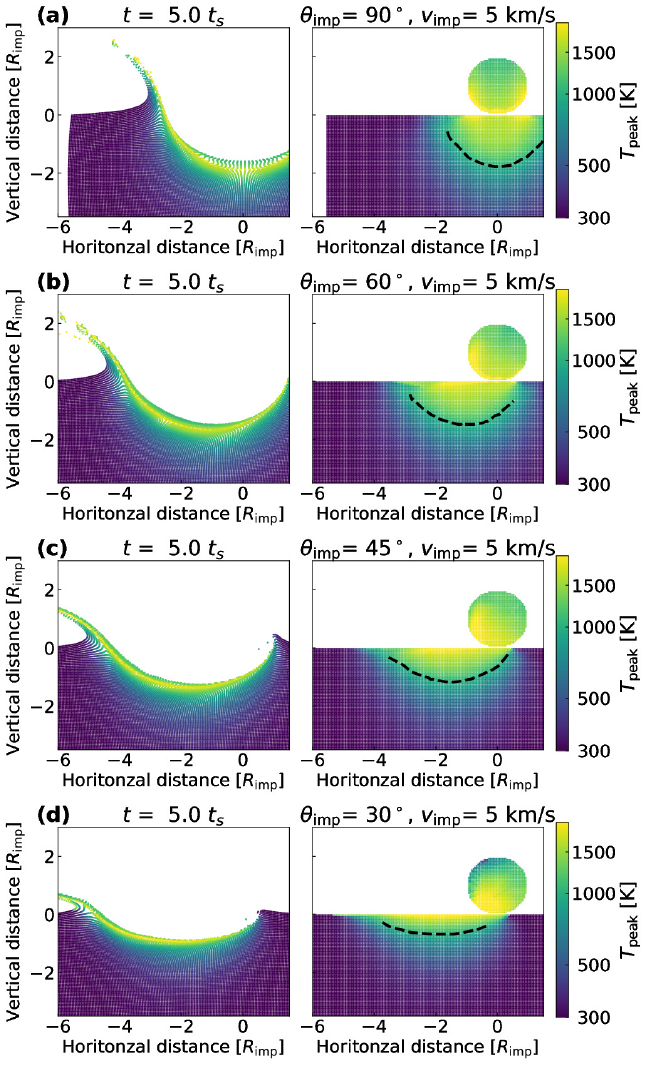}
\caption{
Snapshot of a cross section (x-z plane). 
The impactor hits the target with material strength at (x,z) = (0,0) with $v_{\rm imp}$= 5 km/s and 
(a) $\theta_{\rm imp}$ = 90$^\circ$, (b) $\theta_{\rm imp}$ = 60$^\circ$, (c) $\theta_{\rm imp}$ = 45$^\circ$, and (d) $\theta_{\rm imp}$ = 30$^\circ$respectively.
Each tracer particle is colored by peak temperature ($T_{\rm peak}$).
Left panel depicts time = 5 $t_s$ and 
right panel shows the provenance plots, where we put the tracer particles back in their original locations.
Dashed lines on the right panel indicate the isothermal line of $T_{\rm peak}$= 1000 K.
Note we also plot the tracer particles that are within the overheated regions near the contact boundary between the impactor and the target.
}
\label{fig:material}
\end{figure}

\begin{figure}
\noindent\includegraphics[width=0.9\textwidth]{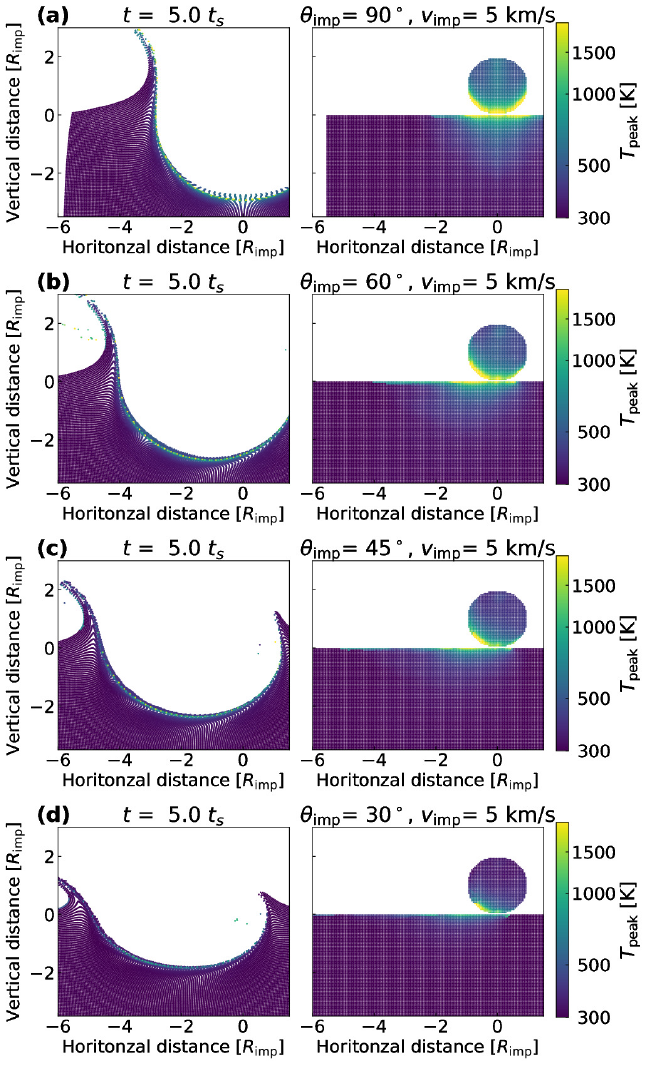}
\caption{Same as Figure \ref{fig:material}, but for the case without material strength.
}
\label{fig:hydro}
\end{figure}

\begin{figure}
\noindent\includegraphics[width=0.9\textwidth]{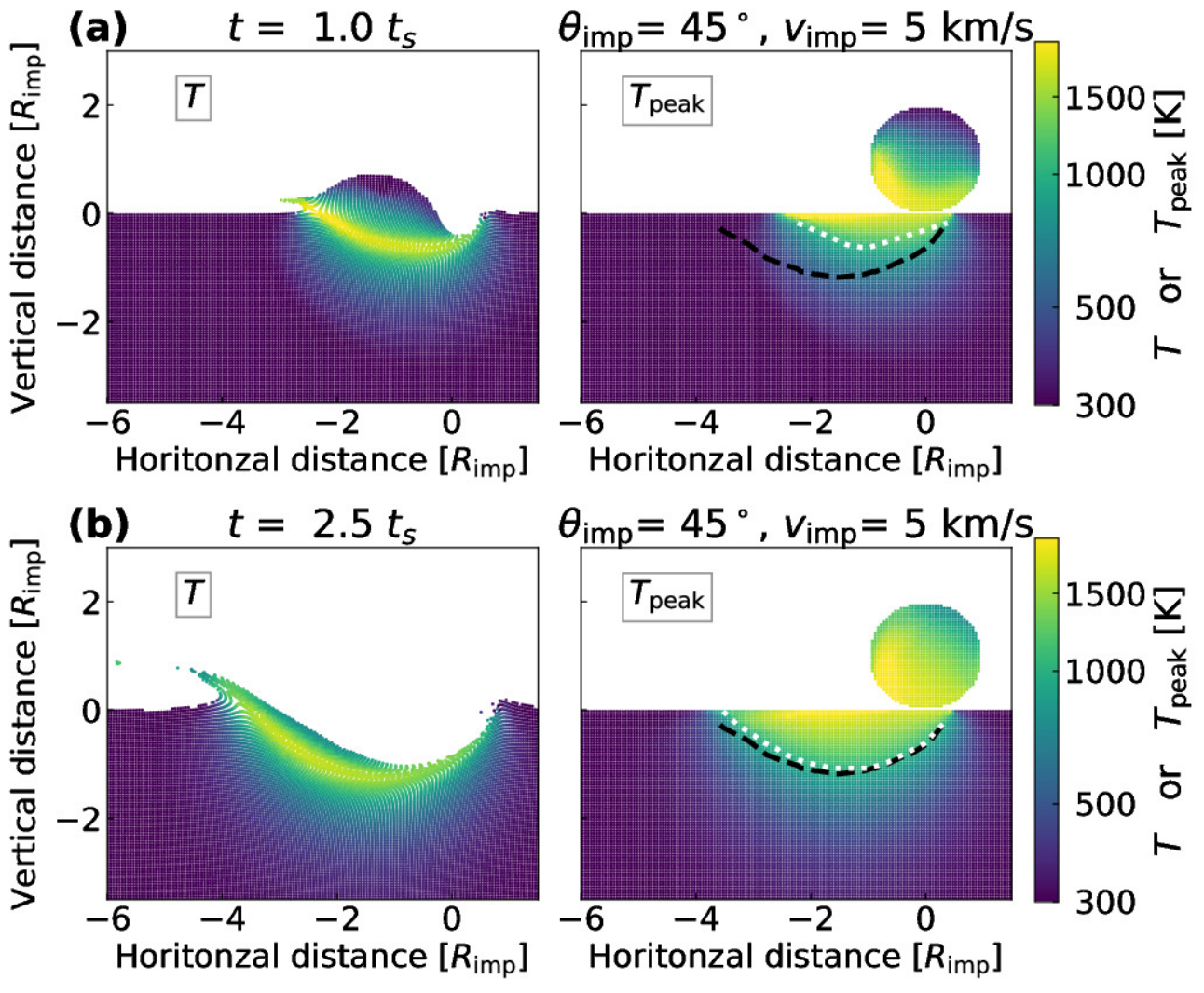}
\caption{
Same as Figure \ref{fig:material} (c) ($\theta_{\rm imp}$ = 45 with $v_imp$ = 5 km/s), but color illustrates temporal temperature ($T$) on left columns and $T_{\rm peak}$ on right columns at (a) 1.0 $t_s$ and (b) 2.5 $t_s$. 
Black dashed lines indicate the isothermal line of $T_{\rm peak}$ = 1000 K at 5 $t_s$ (Figure 1 (c)) and white dotted lines indicate the isothermal line at each time.
}
\label{fig:a45}
\end{figure}

The heated region depends on the impact obliquity.
Figure \ref{fig:material} represents the distribution of peak temperature ($T_{\rm peak}$) at a time of 5 $t_s$ after impact, in a suite of simulations with material strength and an impact velocity of 5~km/s,
where $t_s$ is a characteristic time for impactor penetration, $t_s = 2R_{\rm imp}/v_{\rm imp}$.
As shown in the provenance plots (right panels), the heated area becomes shallower for more oblique impacts (see dashed lines, which show the $T_{\rm peak}=1000$~K isotherm).
Comparing impacts with and without material strength (Figure \ref{fig:hydro}), we can confirm material strength enhances the temperature increase. 
This is consistent with previous work \cite{Kurosawa:2018,Wakita:2019a}.
As confirmed in \citeA{Kurosawa:2018}, more kinetic energy is converted into internal energy in the target with material strength. 
Since deformational heating cannot occur in simulations without material strength, we observe such additional heating only in case with material strength.
Most material has reached its peak temperature at 2.5 $t_s$, as the isothermal line of 1000 K indicates (see white dotted line and black dashed line in Figure \ref{fig:a45} (b)). 
To ensure material has reached its peak temperature we focus our analysis of peak temperatures reached before 5 $t_s$. 
Also, we have confirmed that the cumulative heated mass is similar at 2.5 -- 10 $t_s$ and does not change significantly after 5 $t_s$ \cite<see Figure S5 in>{Wakita:2019a}.

\begin{figure}
\noindent\includegraphics[width=0.9\textwidth]{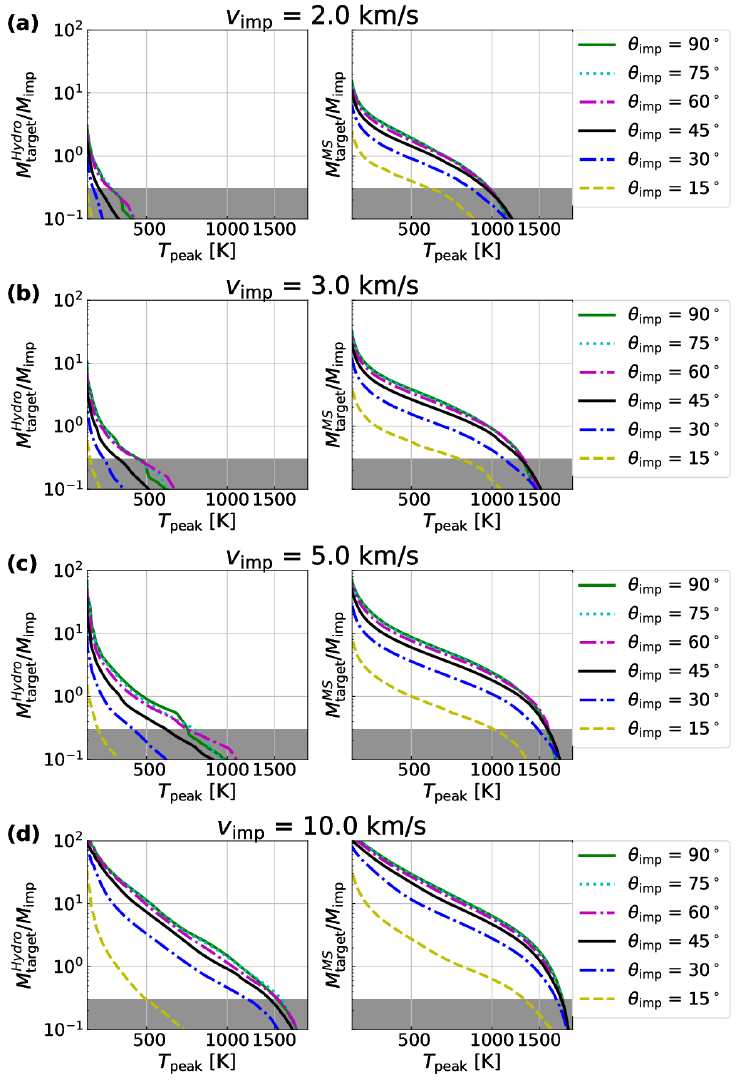}
\caption{Cumulative heated mass of target materials produced by impacts at 
(a) $v_{\rm imp}$ = 2 km/s, 
(b) $v_{\rm imp}$ = 3 km/s, 
(c) $v_{\rm imp}$ = 5 km/s,
and (d) $v_{\rm imp}$ = 10 km/s, respectively.
$M^{\rm Hydro}_{\rm target}$ and $M^{\rm MS}_{\rm target}$ are the cumulative heated mass in the target
without and with material strength, which are normalized by the impactor mass.
Left-hand side in each panel depicts the cases without material strength (the pure hydrodynamic cases) and 
right-hand side with material strength.
Each line represent $\theta_{\rm imp}$ (see legends).
Note that the shaded region depicts the artificial overheated region due to the overshooting in temperature at the contact boundary between the impactor and the target.
}
\label{fig:cummass}
\end{figure}

To examine the effect of material strength, we compare the cumulative heated mass in the target with/without material strength (see Figure \ref{fig:cummass}). 
Note that we consider the number of the tracer particles for which the peak temperature is beyond a given $T_{\rm peak}$, and regard their total mass as the cumulative heated mass.
As shown in Figure \ref{fig:cummass} (c) ($v_{\rm imp}$ = 5 km/s),
the heated mass with material strength (right-hand panel) is always larger than that without material strength (pure hydrodynamic case, left-hand panel) at a given impact angle.
The difference between the case with/without material strength reaches a factor of ten at the most.
Material strength enhances the impact heating regardless of $\theta_{\rm imp}$.
On the other hand, the effect of shear heating for the cumulative heated mass depends on $v_{\rm imp}$.
For lower impact velocity scenarios (Figure \ref{fig:cummass} (a) and (b)), 
our results show the cumulative heated mass in the target with material strength is $\sim$10 times larger than that without material strength.
Previous work indicated a combination of the material strength and movement of the impactor results in the enhanced heating in the oblique impacts \cite{Wakita:2019a}. 
Since the heated mass without material strength at lower velocity has a larger difference between vertical and oblique impacts, it implies that material strength is more effective than a movement of the impactor.
The heated mass without material strength approaches that with material strength as $v_{\rm imp}$ increases (see Figure \ref{fig:cummass} (d)).
This is also consistent with previous findings \cite{Quintana:2015,Kurosawa:2018}.
As a result, the difference between the heated mass of 90$^\circ$ and 45$^\circ$ in the case without material strength approaches to that with material strength (Figure \ref{fig:cummass} (d)).
Thus, the material strength more effectively increases for the cumulative heated mass in lower $v_{\rm imp}$ scenarios than higher $v_{\rm imp}$.

We now consider the results with material strength and discuss the effect of $\theta_{\rm imp}$ on impact heating.
Previous work focused on oblique impacts at $v_{\rm imp}$ = 5 km/s showed that the cumulative heated mass of 90$^\circ$ and 45$^\circ$ are almost the same \cite{Wakita:2019a}.
We find the heated mass of 75$^\circ$ and 60$^\circ$ are between that of 90$^\circ$ and 45$^\circ$, and their difference is within a factor of 1.5 (Figure \ref{fig:cummass} (c)).
On the contrary, we find that shallower impacts ($\theta_{\rm imp} \le 30^\circ$) produce much less heated mass than 45$^\circ$ impacts.
Grazing impactors are decapitated before they penetrate into the target and heating is limited mainly to the lower hemisphere \cite{Schultz:1990,Davison:2011}.
While the
heated area of $\theta_{\rm imp} \le 30^\circ$ impacts has a similar width as the 45$^\circ$ impacts,
the shallower penetration of grazing impactors results in the heated area extending to smaller depths 
(see dashed lines in Figure \ref{fig:material} (c) and (d)).
As a result, the heated mass of $\theta_{\rm imp} \le 30^\circ$ becomes smaller than other higher angle impacts.
Note that it is beyond the focus of our paper to find the threshold impact angle where cumulative heated mass begins decreasing.

While the cumulative heated mass takes a similar value for $>$ 45$^\circ$ impacts,
the cumulative heated mass by $<$ 30$^\circ$ impacts is always less than that regardless of the impact velocities ($v_{\rm imp}$).
Our results show impacts of $\theta_{\rm imp} \ge 45^\circ$ produce a similar heated mass within less than a factor of 1.5 (right panels in Figure \ref{fig:cummass}). 
Note that the results saturates around $M_{\rm target}^{\rm MS}/M_{\rm imp} \sim 100$ (see Figure \ref {fig:cummass} (d)),
because our total computational region in the target that we track with tracer particles is as large as $M_{\rm target}/M_{\rm imp} \simeq 10^2 $ (see Methods).
On the contrary, the ratio of the cumulative heated mass with shallower angles ($\theta_{\rm imp} \le 30^\circ$) to that with $\theta_{\rm imp}=45^\circ$ ($M_{\rm target}^{\rm MS} (\theta_{\rm imp})/M_{\rm target}^{\rm MS} (\theta_{\rm imp} = 45^\circ))$ is $\sim$ 0.6 ($\theta_{\rm imp}=30^\circ$) and $\sim$ 0.2 ($\theta_{\rm imp}=15^\circ$).
Nevertheless,
the heated mass from shallower angle impacts ($\theta_{\rm imp} \le 30^\circ$) is less than steeper angle impacts, 
the effect of material strength on the degree of impact heating is still significant (see Figure \ref{fig:cummass}).

\section{Discussion}
\begin{figure}
\noindent\includegraphics[width=\textwidth]{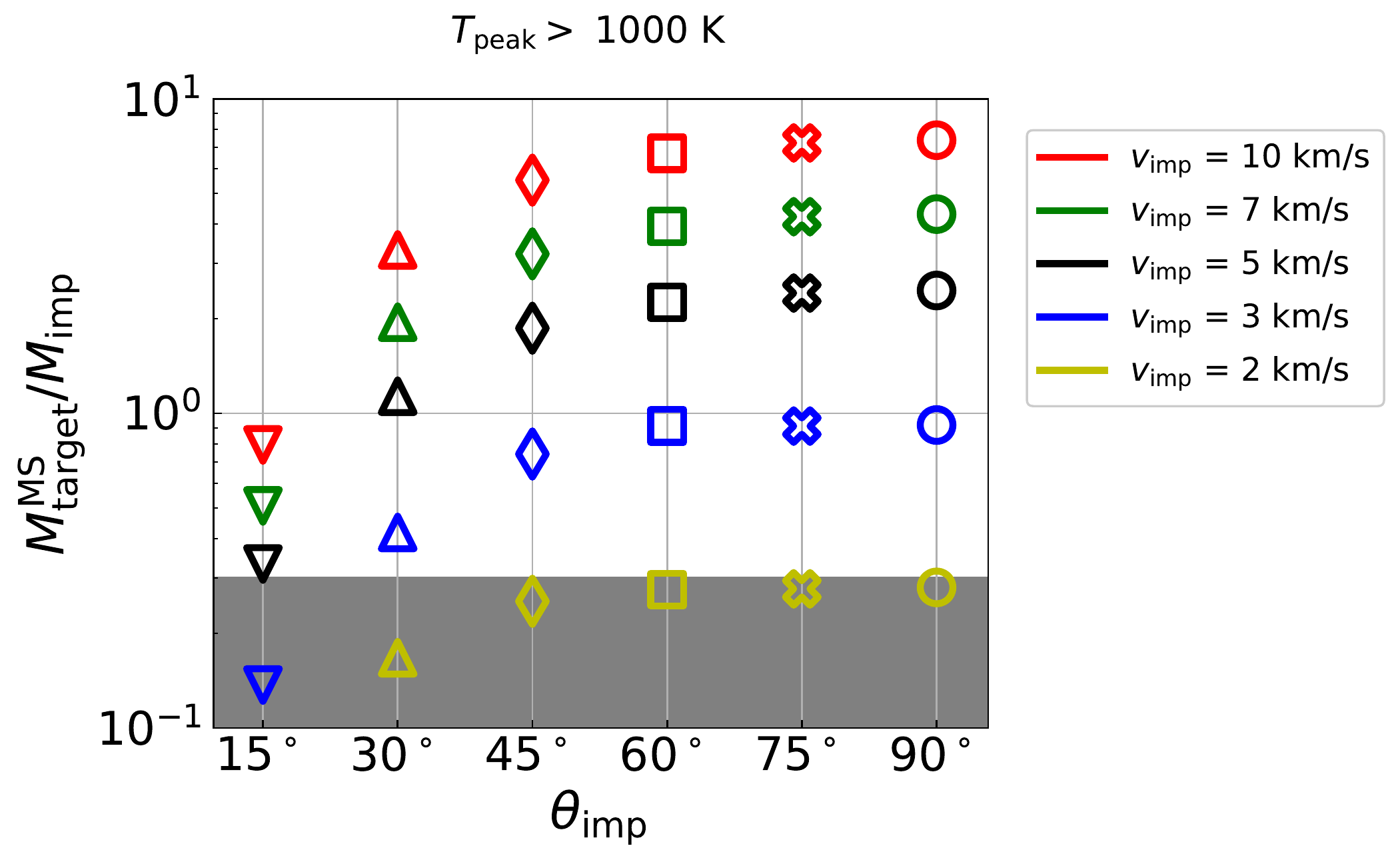}
\caption{Cumulative heated mass of  ${T}_{\rm peak}$ over 1000 K in the target with material strength as a function of $\theta_{\rm imp}$. 
Symbols are colored according to $v_{\rm imp}$ (see legend).
The shaded region indicates the artificial overheated region near the contact of the impactor and target.
}
\label{fig:tpeak}
\end{figure}

Here we discuss the impact induced heated materials using our results.
If the impact heats the target enough beyond a threshold temperature, it could trigger Ar loss and reset the target's $^{40}$Ar-$^{39}$Ar age.
\citeA{Kurosawa:2018} showed that the threshold of impact velocity for Ar loss would be 2 km/s in the target with the material strength,
which is lower than 8 km/s in the case without material strength.
We investigate the cumulative heated mass of the impact-induced Ar age resetting during oblique impacts,
by assuming 1000 K as an index temperature, which is the closure temperature of typical Ar carrier mineral \cite<e.g., feldspar,>{Cohen:2013}.
Figure \ref{fig:tpeak} summarizes the cumulative heated mass of $T_{\rm peak} > 1000$ K in case of the material strength.
The dependence of the cumulative heated mass on $\theta_{\rm imp}$ are similar regardless of $v_{\rm imp}$;
the heated mass of $\theta_{\rm imp} \ge 45^\circ$ are always larger than $\theta_{\rm imp} \le 30^\circ$ at given $v_{\rm imp}$.
The high-velocity impacts produce a larger amount of the cumulative heated mass than the low-velocity case.
While $^{40}$Ar-$^{39}$Ar age has been used to estimate the latest impact event on the parent body of meteorites \cite{Bogard:2011}, it is also possible to estimate their original depth. 
When $M^{\rm MS}_{\rm target}/M_{\rm imp}$ is sufficiently large (e.g., $\ge 1$), we can regard such an impact condition as resetting the Ar age.
Thus, our results showed that grazing impacts (e.g., $\theta_{\rm imp} = 30^\circ$ with 5 few km/s) can contribute to the Ar age resetting (Figure \ref{fig:tpeak}). 
Since grazing impacts excavate relatively shallower material (see Figure \ref{fig:material} (d)), it may imply that meteorites might originate from shallower depth than previously thought. 

\begin{figure}
\noindent\includegraphics[width=\textwidth]{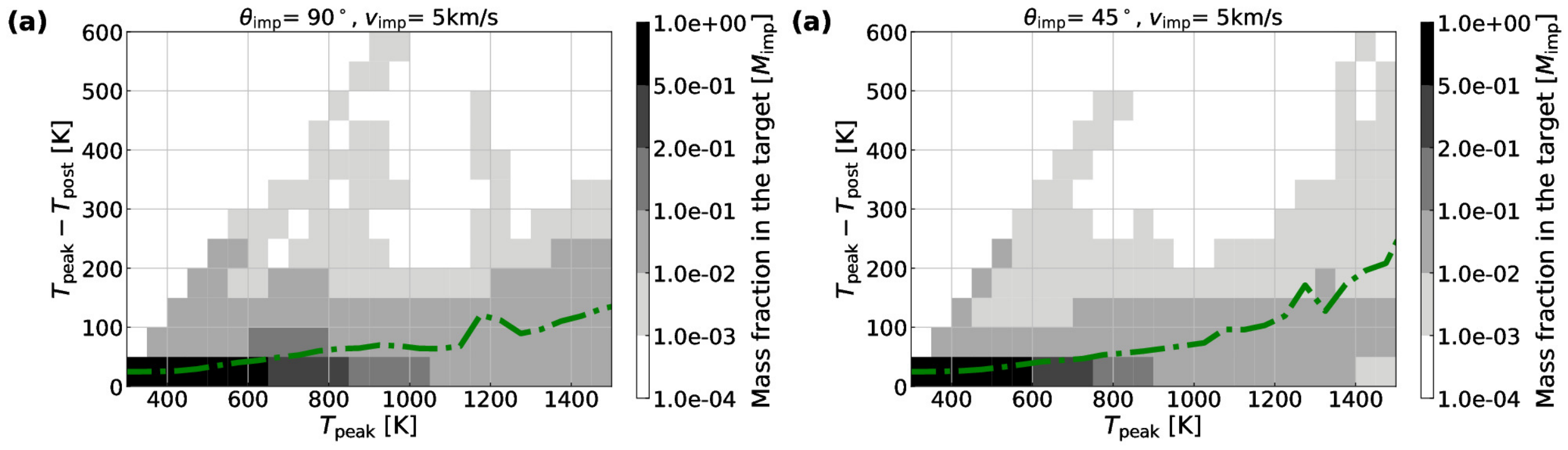}
\caption{Heatmaps of difference between $T_{\rm peak}$ and post temperature $T_{\rm post}$ at the time of 5 $t_s$,
(a) $\theta_{\rm imp}$ = 90$^\circ$ with $v_{\rm imp}=5$ km/s and (b) $\theta_{\rm imp}$ = 45$^\circ$ with $v_{\rm imp}=5$ km/s.
The gray contour represents their mass fraction in the target normalized by the mass of the impactor.
The cell size is 50 K; for example, the cell of ($T_{\rm peak}$ = 1000--1050 K, $T_{\rm peak}$ - $T_{\rm post}$ = 50--100K) represents the mass fraction is on the order of $10^{-2}$ for both cases. 
A green dash-dotted line indicates the weighted average of $T_{\rm peak} - T_{\rm post}$.
}
\label{fig:diff}
\end{figure}

\begin{figure}
\noindent\includegraphics[width=\textwidth]{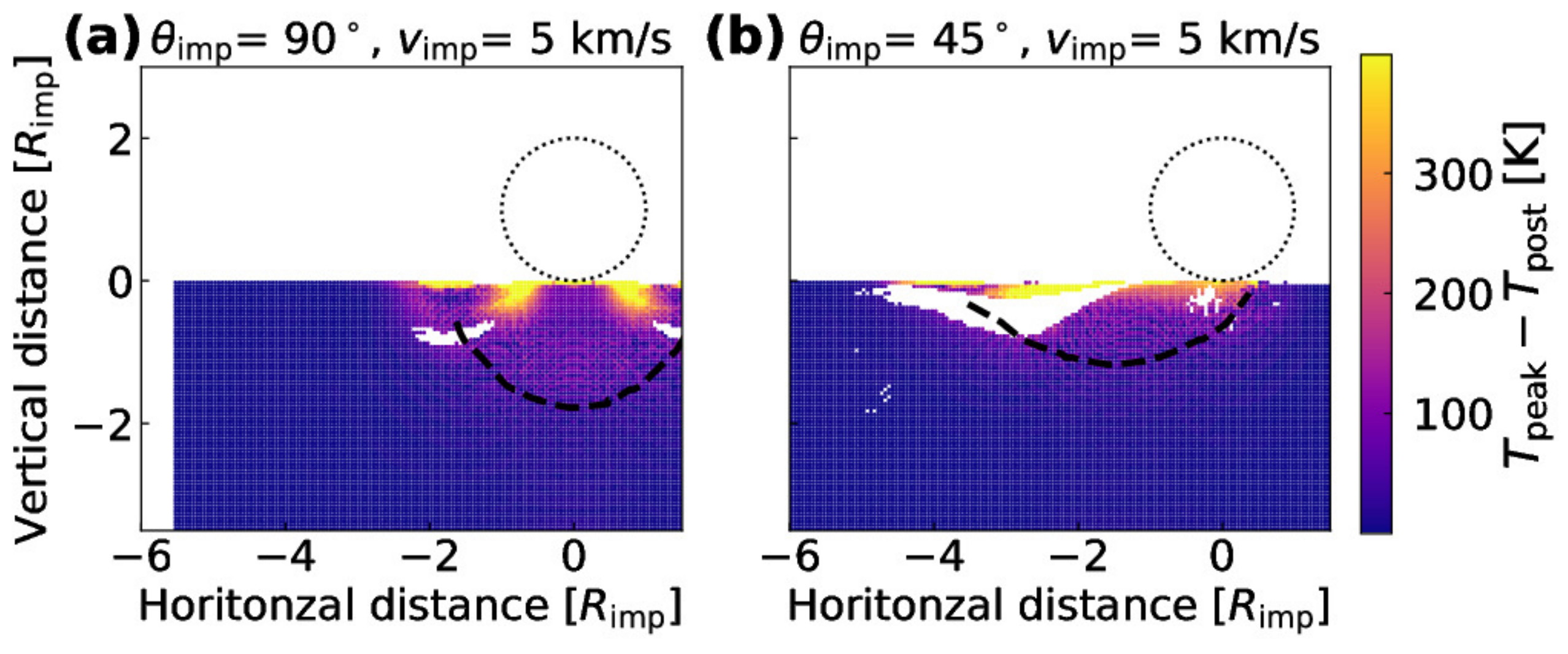}
\caption{
Same viewing in right columns of Figure \ref{fig:material}, but color illustrates $T_{\rm peak}$ - $T_{\rm post}$ at 5 $t_s$, (a) $\theta_{\rm imp}$ = 90$^\circ$ with $v_{\rm imp}$ = 5 km/s and (b) $\theta_{\rm imp}$ = 45$^\circ$ with $v_{\rm imp}$ = 5 km/s. 
Note that we only plot the tracer particles in the target that decrease to $10^5$ Pa within 5 $t_s$ (white regions are out of this condition). 
Dashed lines indicate the isothermal line of $T_{\rm peak}$ = 1000 K in Figure \ref{fig:material} (a) and (c). 
Dotted circles represent the provenance location of the impactor.
}
\label{fig:tdiff_plot}
\end{figure}

We need to mention that the usage of $T_{\rm peak}$ can exaggerate the cumulative heated mass.
Because the $T_{\rm peak}$ records the temperature during the shock, it may be overestimated due to the numerical diffusion. 
Even if the $T_{\rm peak}$ is the same, the difference in pressure may indicate the different status (i.e., shock state or post-shock state). 
In such a case, especially for the phenomena that would take a time to occur (e.g., dehydration), the post-shock temperature ($T_{\rm post}$, the temperature after the pressure becomes less than $10^5$ Pa) can be useful. 
When we compare the peak temperature with the post-shock temperature, 
we find that the former is overestimated by about 100 K in comparison with the latter at $T_{\rm peak}=$ 1000 K in the case of $v_{\rm imp}=5$ km/s (Figure \ref{fig:diff}).
Although $T_{\rm post}$ is a more accurate way, it is computationally expensive to examine all tracer particles and incapable in our current work. 
Because some particles, that are at the contact between the impactor and the target, take a longer time ($> 5 t_s$) to decrease pressure to $10^5$ Pa (see white region in Figure \ref{fig:tdiff_plot}). 
However, the difference between $T_{\rm peak}$ and $T_{\rm post}$ are acceptable and our estimate on $T_{\rm peak}$ is appropriate in accounting the cumulative heated mass. 
\citeA{Kurosawa:2018} calculated the cumulative heated mass for Ar loss using the entropy corresponding to 1000 K at $10^5$ Pa.
Our result of $M^{\rm MS}_{\rm target}/M_{\rm imp} = 7.04$ with a vertical impacts at $v_{\rm imp}=$ 10 km/s 
are within $11\%$ difference from their result of 6.32 \cite<see Figure 3 in>{Kurosawa:2018}.
This agreement, suggests that the use of peak temperature to estimate the cumulative heated mass after the impact-induced shock heating is reasonable.

\begin{figure}
\noindent\includegraphics[width=\textwidth]{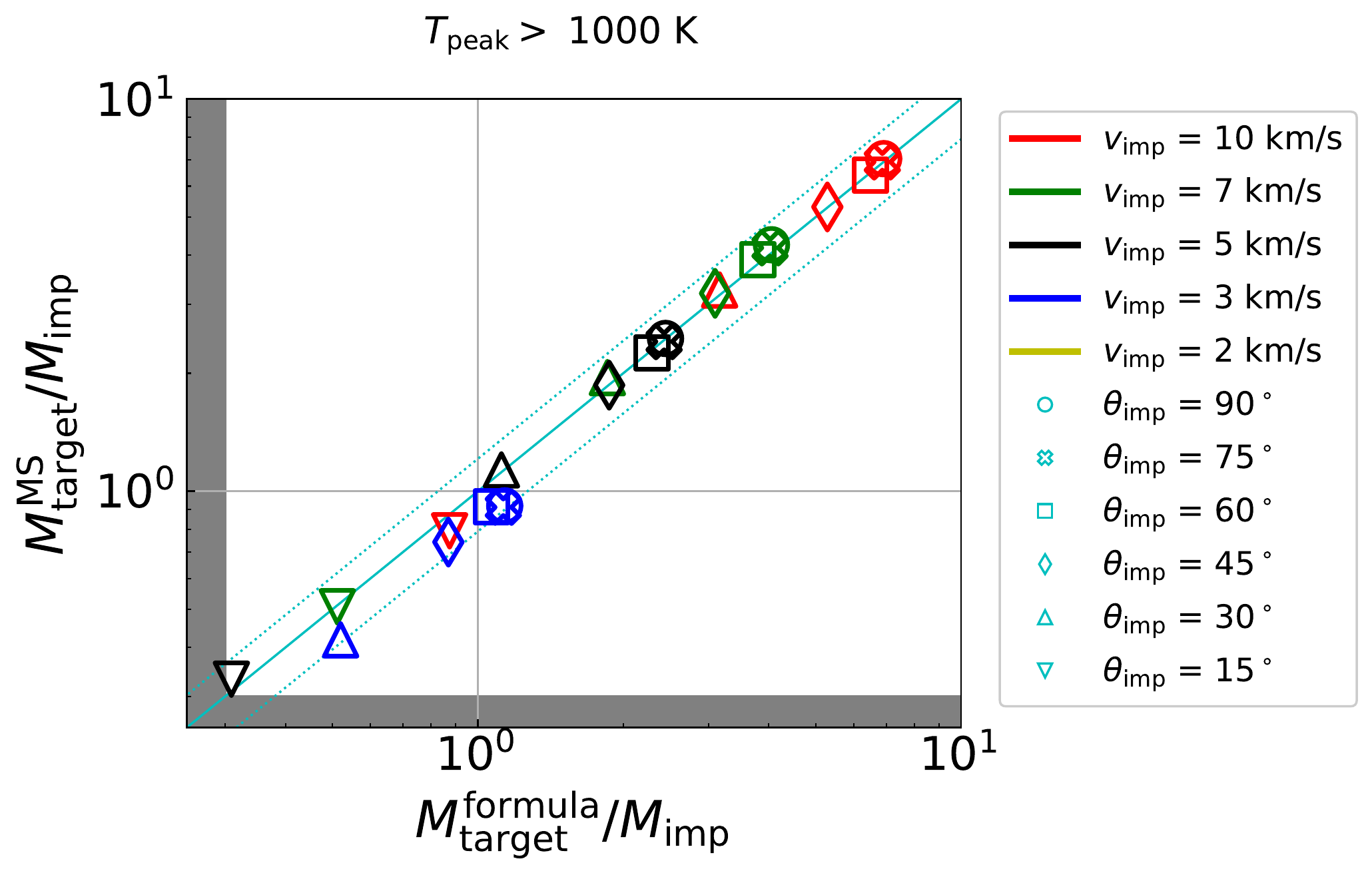}
\caption{Comparison of cumulative heated mass in the target ($T_{\rm peak}$ $>$ 1000 K). 
While numerical results are plotted on y-axis, the results from the empirical formula (Equation \ref{eq:MS}) are shown in x-axis.
The gray solid diagonal line indicates both results are the same, and the dotted lines represent they are within an error of $\pm 21\%$.
The symbols are the same viewing as in Figure \ref{fig:tpeak} (see legend).
}
\label{fig:tpeak_comp}
\end{figure}

It is worth exploring the dependence of the cumulative heated mass on impact angle and velocity. 
To estimate the cumulative heated mass at various impact velocities and angles, we derive an empirical formula.
Note that an artificial increase in temperature around the contact boundary between impactor and target in numerical simulations is reported \cite{Kurosawa:2018}.
We use the numerical results over 0.3 $M_{\rm imp}$ which are free from the artificial overheating and thus more conservative. 
To minimize the number of coefficients in the empirical formula, we adopt the following equation, 
\begin{linenomath*}
\begin{equation}
M^{\rm formula}_{\rm target} (T_{\rm peak})/M_{\rm imp} = 
\left\{ C_1 \sin(\theta_{\rm imp}) + C_2 \sin^2(\theta_{\rm imp}) + (1-C_1-C_2)\sin^3(\theta_{\rm imp}) \right\} C_3(0.5 v^2_{\rm imp}/E_{T_{\rm peak}})^{C_4},
\label{eq:MS}
\end{equation}
\end{linenomath*}
where $E_{T_{\rm peak}}$ is the specific internal energy at $T_{\rm peak}$.
The procedure to obtain the empirical formula is described as follows. 
We prepare five datasets at a given impact velocity (2, 3, 5, 7, and 10 km/s), which are the normalized heated mass as a function of impact angle. 
We confirm that the five datasets almost converge into a single trend against the change in impact angle. 
Then, we fit the all data with a third order polynomial function with two physical constraints, 
which are (1) the normalized heated mass equals zero at $\theta_{\rm imp} = 0^\circ$, 
and (2) the normalized heated mass equals unity at $\theta_{\rm imp} = 90^\circ$. 
Thus, the dependence of impact angle on the heated mass at a given impact velocity can be described with two fitting coefficients ($C_1$ and $C_2$ within bracket in Equation \ref{eq:MS}). Next, we divide the heated masses by the effects of the impact angle, resulting in the corrected heated mass depending only on impact velocity, which corresponds to the specific internal energy. 
The corrected data points can be fitted well by a power-law with two fitting constants ($C_3$ and $C_4$ in Equation \ref{eq:MS}). By combining impact angle and velocity dependence, we obtain the empirical formula shown as Equation (\ref{eq:MS}) with four coefficients.
For the case of $T_{\rm peak}=$ 1000 K (Figure \ref{fig:tpeak}), we find that the coefficients are $C_1=-0.249$, $C_2=3.40$, $C_3=1.07$, and $C_4=0.749$, respectively.
Note that $E_{T_{\rm peak}} = 4.16$ MJ/kg at $T_{\rm peak}=$ 1000 K. 
Figure \ref{fig:tpeak_comp} represents that the empirical formula works properly within the error of $\pm 21\% (2 \sigma)$  to our numerical results with material strength.
This formula would be useful to estimate the cumulative heated mass resulting from various impact conditions. 

\begin{figure}
\noindent\includegraphics[width=\textwidth]{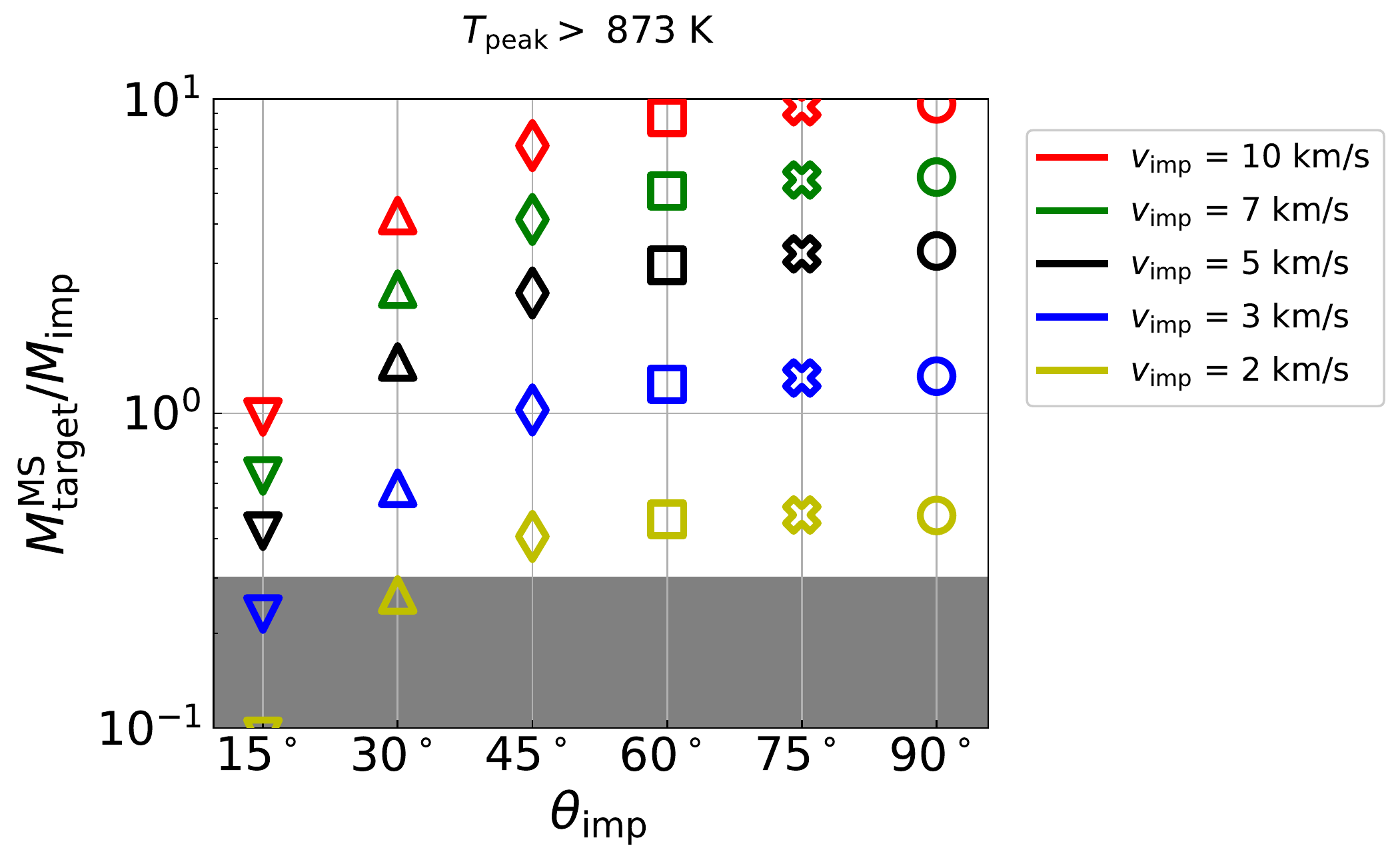}
\caption{Same as Figure \ref{fig:tpeak}, but for the case of $T_{\rm peak}>873$ K.
}
\label{fig:tpeak_873K}
\end{figure}

\begin{figure}
\noindent\includegraphics[width=\textwidth]{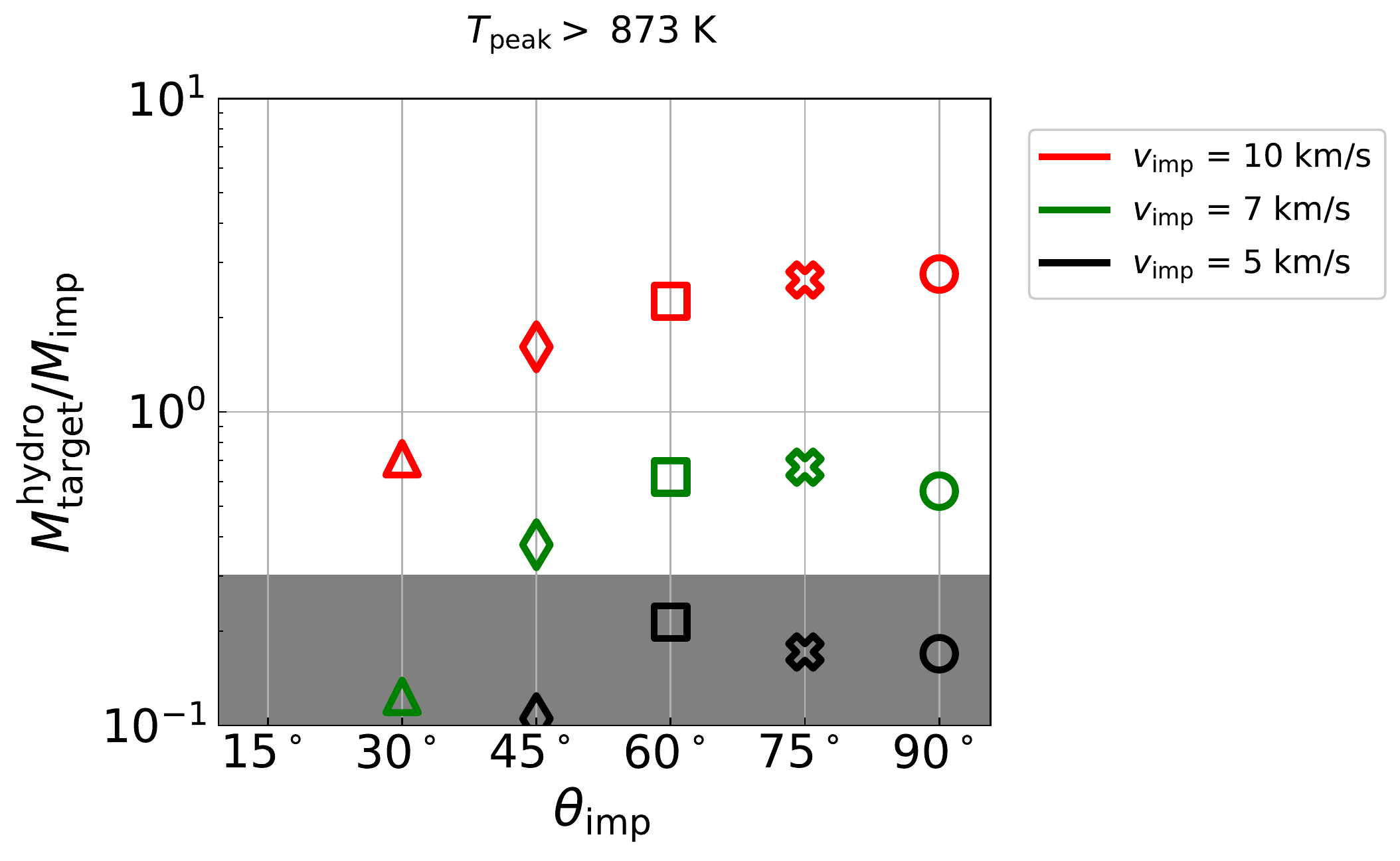}
\caption{Same as Figure \ref{fig:tpeak_873K}, but for the case without material strength.
}
\label{fig:tpeak_873Khydro}
\end{figure}

\begin{figure}
\noindent\includegraphics[width=1.2\textwidth]{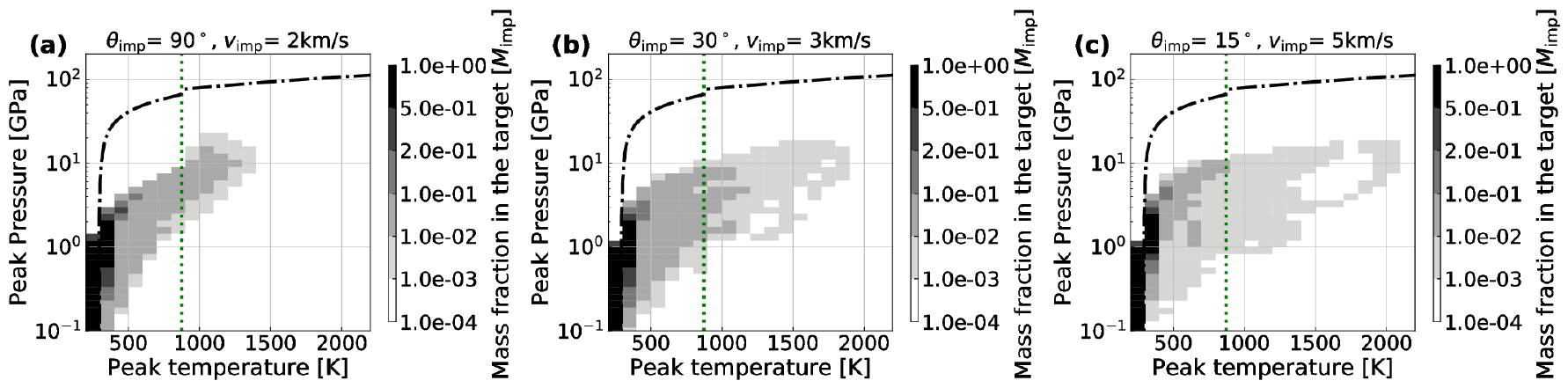}
\caption{Heatmaps of peak pressure and peak temperature at the time of 5 $t_s$,
(a) $\theta_{\rm imp}$ = 90$^\circ$ with $v_{\rm imp}=2$ km/s, (b) $\theta_{\rm imp}$ = 30$^\circ$ with $v_{\rm imp}=3$ km/s,
and (c) $\theta_{\rm imp}$ = 15$^\circ$ with $v_{\rm imp}=5$ km/s, respectively.
The green dotted vertical line indicates $T_{\rm peak}$= 873 K, the temperature threshold for the dehydration.
The black dot-dashed line represents the Hugoniot curve of dunite.
The gray contour represents the mass fraction in the target normalized by the mass of the impactor.
}
\label{fig:ppeak_tpeak}
\end{figure}

We also discuss the occurrence of dehydration reactions of phyllosilicate (e.g., serpentine).
As the dehydration of hydrous materials is an endothermic reaction, additional calculations are required to assess the dehydrated mass accurately \cite<e.g.,>{Kurosawa:2021a}.
Additionally, \citeA{Kurosawa:2021a} experimentally showed that the efficiency of impact devolatilization on carbonaceous asteroid-like materials (e.g., calcite) is considerably low ($<10\%$ of theoretical prediction).
Thus, the dehydrated mass estimated by numerical simulations will overestimate regardless of considering the reaction heat.
Although we may overestimate the heated mass by using peak temperature but ignoring the reaction heat, it is worth considering the impact conditions that may induce the dehydration.
We here take a temperature threshold for the dehydration as 873 K, 
at which the dehydration of phyllosilicate starts to occur \cite{Lange:1982,Nozaki:2006,Nakato:2008}.
Note that the Hugoniot curve for dunite and serpentine differs \cite{Benz:1989,Brookshaw:1998}.
Based on shock heating alone, the serpentine will reach a similar temperature of the dunite at a given impact condition. 
Thus, the amount of dehydrated mass may not change even if the equation state of serpentine is used.
Please also note that our numerical setup is different from previous work of investigating the fate of hydrous minerals during the impacts which suggested that the serpentine in the core can avoid the dehydration \cite{Wakita:2019}.
Since they considered a dunite layer over the serpentine core, the dunite layer might insulate the serpentine core from the direct impact-induced heating that we explore in this work.
The vertical impact with material strength at $v_{\rm imp}$ = 2 km/s produces dehydrated materials of $M^{\rm MS}_{\rm target}/M_{\rm imp} \sim 0.4$ (Figures \ref{fig:cummass} and \ref{fig:tpeak_873K}). 
Grazing impacts  ($\theta_{\rm imp} \le 30^\circ$) into the target with material strength require a higher impact velocity to produce a similar amount of impact heated material as vertical impacts (see also Figure \ref{fig:tpeak_873K});
$\theta_{\rm imp} = 30^\circ$ with $v_{\rm imp}$ = 3 km/s ($M^{\rm MS}_{\rm target}/M_{\rm imp} \sim 0.5$)
and $\theta_{\rm imp} = 15^\circ$ with $v_{\rm imp}$ = 5 km/s ($M^{\rm MS}_{\rm target}/M_{\rm imp} \sim 0.4$). 
Nevertheless, those grazing impacts with material strength require lower velocities than vertical impacts without material strength to produce a similar amount of impact heated material
(at an impact velocity of $v_{\rm imp}$ = 7 km/s, $M^{\rm Hydro}_{\rm target}/M_{\rm imp} \sim 0.5$, Figure \ref{fig:tpeak_873Khydro}).
Oblique impacts generate heated materials at lower peak pressure than vertical impacts at a given impact velocity \cite{Wakita:2019a}.
We also find grazing impacts at higher impact velocities could produce dehydrated material at lower peak pressures than vertical impacts with lower impact velocity (see Figure \ref{fig:ppeak_tpeak}).
As the high temperature region of oblique impacts is widely distributed (see Figure \ref{fig:material}), the weakly shocked region is also close to the surface, potentially to be ejected. 
If this ejected material eventually lands on the Earth as meteorites, this implies that shock heated meteorites could have experienced a wide range of peak pressures.
Thus, oblique impacts of $\theta_{\rm imp} < 45^\circ $ may have produced dehydrated minerals in weakly shock-metamorphosed meteorites.

The time takes for hydrous minerals to dehydrate depends on the reaction temperature.
While dehydrated materials in carbonaceous chondrites indicate shock heating \cite<e.g.,>{Nakamura:2005,Nakato:2008,Abreu:2013}, 
carbonaceous chondrites have generally experienced weak shocks.
The dehydration starts to occur at about 873 K \cite<600 $^\circ$C,>{Lange:1982,Nozaki:2006,Nakato:2008}, but some work examines the duration time at higher temperature.
\citeA{Nozaki:2006} conducted heating experiments on carbonaceous chondrites: 
Both short (10 s) and long (120 s) heating at a temperature of 1173 K (900 $^\circ$C) decomposed the hydrous minerals. 
They indicated that the temperature is more important for dehydration than duration. 
Other experimental work implies that a weakly-shocked and dehydrated carbonaceous chondrite would have experienced 1--100 hour heating at 1173 K or a 10-1000 days heating at 973 K (700 $^\circ$C) \cite{Nakato:2008}. 
While our impact simulations are unable to consider the duration materials spend at elevated temperatures, we can estimate the cooling time of the hydrous materials. 
Assuming that hydrous material is heated to a depth of 10 m ($d$, smaller than our cell size), its cooling time scale would be over 10,000 days. 
Note that we take the thermal diffusivity ($\kappa$) of $10^{-7} {\rm m}^2/{\rm s}$, which is a typical value of the carbonaceous chondrite \cite{Opeil:2020}, and calculate $t_{\rm cool} \sim d^2/\kappa$. 
The material in much deeper locations would take longer time.
Although we are unaware of the time to dehydrate at 873 K, the cumulative heated mass above 873 K may represent the amount of dehydrated minerals. 

\section{Conclusions}
The mass heated by oblique impacts is a key to the understanding of the history of asteroids and meteorites. 
While \citeA{Kurosawa:2018} clarified the importance of the material strength of the target in the vertical impact, the following work confirmed this in the oblique impacts at a given impact velocity and angles \cite{Wakita:2019a}. 
We advanced their work by examining the dependence of the impact velocities and angles.
Considering material strength in the target, we have performed a series of oblique impact simulations over a range of impact velocities and angles.
The cumulative heated mass in the target with material strength is always larger than that without material strength,
which indicates the material strength enhances impact heating.
Our oblique impacts simulations with material strength showed that vertical impacts and impacts with steeper angles ($\geq$ 45$^\circ$) 
generate a similar cumulative heated mass, within a factor of 1.5.
Grazing angle impacts ($\leq$ 30$^\circ$) produce less heated mass than other oblique impacts regardless of impact velocity.
From our impact simulations of a wide parameter space, 
we derived an empirical formula for material with peak temperature over 1000 K, which can be used to understand  $^{40}$Ar-$^{39}$Ar age resetting. 
Vertical impacts at low impact velocity and grazing impacts at high impact velocity produce a similar heated mass, 
but have differences in their peak pressure, indicating that grazing impacts are more likely to be responsible for impact heating in weakly shocked meteorites.

%
%
%
%
%
%
%
%

\section*{Open Research}
All our data are given by using iSALE-3D and our input files are available \cite{Wakita:2022a}.
We also provide the cumulative heated mass of various impacts with/without material strength for $T_{\rm peak}$ of every 10 K as a Data Set.
Please note that usage of the iSALE-3D code is restricted to those who have contributed to the development of iSALE-2D, 
and iSALE-2D is distributed on a case-by-case basis to academic users in the impact community.
It requires a registration from the iSALE webpage (https://isale-code.github.io/) and usage of iSALE-2D and computational requirements are also shown in there.  
We directly plot figures from our binary data using pySALEPlot which is included in iSALE-3D and developed by TMD. 
Please also note that pySALEPlot in the current stable release of iSALE-2D (Dellen) would not work for the data from iSALE-3D.

\acknowledgments
We gratefully acknowledge the developers of iSALE-3D, including Dirk Elbeshausen, Kai W{\"u}nnemann, and Gareth Collins. 
This work has been supported in part by JSPS, Japan KAKENHI Grant Number JP17H06457 and JP17H02990.
HG and KK are supported by JSPS KAKENHI Grant JP19H00726.
KK is supported by JSPS KAKENHI Grants JP18H04464 and JP21K18660. 
TMD is funded by STFC grant ST/S000615/1.


%
%


%
%
%
%
%

\end{document}